\documentclass[11pt, oneside]{article}   	
\usepackage{graphicx}				
\usepackage{cite}
\usepackage{amssymb}

\title{Aspects of NMR Reciprocity and Applications in Highly Conductive Media}

\author{Andrew J. Ilott, and Alexej Jerschow*\\Department of Chemistry\\New York University\\ New York, NY 10003\\\vspace{1cm}\\ * corresponding author: alexej.jerschow@nyu.edu}

\begin{document}
\maketitle

\section*{Abstract}
In the context of NMR spectroscopy and MRI, the principle of reciprocity provides a convenient method for determining the reception sensitivity from the transmitted rf field pattern. The reciprocity principle for NMR was originally described by Hoult et al. [J. Magn. Reson. (1976),24, 71], and can be seen as being based on the broader Lorentz reciprocity principle, and similar theorems from antenna theory. One frequent application of the reciprocity principle is that for a single coil used for both transmission and detection, the transmit and receive fields can be assumed to be equal. This aspect is also where some of the conceptual difficulty of applying the theorem may be encountered. For example, the questions of whether one should use the complex conjugate field for detection, or whether one should apply the theorem in the rotating frame or the laboratory frame are often where considerable confusion may arise, and incorrect results may be derived. We attempt here to provide a helpful discussion of the application of the reciprocity principle in such a way as to clarify some of the confounding questions. In particular, we avoid the use of the `negatively rotating frame', which is frequently mentioned in this context, since we consider it to unnecessarily complicate the matter. In addition, we also discuss the implications of the theorem for magnetic resonance experiments on conducting samples, and metals, in particular.

\section{Introduction}



The principle of reciprocity for magnetic resonance \cite{hoult1976signal,hoult2011signal} has been used to determine coil sensitivities and the spatial distributions of the detected signals from the knowledge of the transmitted field values.  In the description provided below, we lean heavily on concepts discussed in Refs \cite{brown2014magnetic,hoult2000principle,keun2014electro,sodickson2018optimality}, but provide a somewhat different account of the matter in an attempt to clarify certain obscur points and provide a straightforward conceptual picture of the effects in question. As mentioned by Hoult \cite{hoult2000principle}, complex numbers are often used in two different ways in the description of the fields, from which inconsistencies can arise. One of their uses is to indicate the sense of rotation (positively and negatively rotating frames), while the other is used to refer to time lags of fields and signals in the rotating frame \cite{hoult2000principle}. Here, we will use complex numbers exclusively for time lags, and avoid, in particular, the discussion of the negatively rotating frame, as we believe such a construct is more of an impediment than help in properly applying the reciprocity principle. The purpose of this article is not so much the derivation of the reciprocity principle, which is well known, but rather  its application, since this is where problems often arise. For completeness, a derivation is also listed in the Appendix.
  
One particularly illustrative demonstration of the reciprocity principle consists of describing an experiment wherein one uses two coils, with one being driven by an ac or rf field and the other one connected to a detector/oscilloscope \cite{hoult2000principle}. The voltage induced in one coil by driving the other with unit current is the same as when the roles of the coils are reversed. This outcome is found irrespective of the shapes and the geometrical arrangements of the two coils. By extension, it was argued, one could think of the effect of the driving coil as being represented by a magnetic dipole, or a magnetization $\mathbf{M}$ corresponding to the induced field that would be created by the current passing through the coil (see Fig. 1). For example, for a small circular loop, the strength of the dipole would be given by $\mathbf{\mu} = I a$, where $I$ is the current, and $a$ is the area of the loop. For irregular shapes, the use of the magnetization $\mathbf{M}$,  would be a more suitable approach, since $\mathbf{M}$ describes the spatially varying dipole density and thus provides a more general description. One could thus think of an irregularly-shaped coil, which, when a current passes through it, is modeled by a spatially varying $\mathbf{M}$ distribution. A particularly visual illustration of these points is made in Fig. 2 of Ref \cite{hoult2011principle}.   

The electromotive force (emf) $\mathcal{E}$ induced in the receiving coil is given by 
\begin{equation}
	\mathcal{E}=-d\Phi/dt,
\end{equation}
where $\Phi$ is the flux through the coil determined by
\begin{equation}
  \Phi = \int_{A} \mathbf{B^M} \cdot d \mathbf{a}.
\end{equation}
$\mathbf{B^M}$ is the field produced by the magnetization, and the integral runs over the receiving coil area.  

The central NMR reciprocity expression is then 
\begin{equation}
	\mathcal{E} = -\frac{d}{dt}   \int_V \mathbf{B}_1(\mathbf{r}) \cdot \mathbf{M}(\mathbf{r},t) \/ d^3 r, \label{eq:reciproc}
\end{equation}
where the integration is carried out over the sample volume $V$, $\mathbf{M}$ is the sample magnetization, and $\mathbf{B}_1$ is the rf field generated by the receive coil at the position of the magnetization if a unit current were passed through it \cite{hoult2000principle,hoult2011principle}. Hoult \cite{hoult2011principle} has described the outlines of a straightforward derivation of this relationship. 

Eq. (\ref{eq:reciproc}) is to be interpreted as follows: A unit magnetization at  position $\mathbf{r}$ will induce an emf in the receiving coil. This emf is proportional to the magnetic field $\mathbf{B}_1$ that would be generated by the receiving coil at position $\textbf{r}$ if a unit current were passed through the coil. 

Considering that a derivative with respect to time in Eq. (\ref{eq:reciproc}) needs to be carried out, it is convenient to express all relevant quantities in time-harmonic form, 
that is
\begin{eqnarray}
   \mathbf{B}_{1,lab}(\mathbf{r},t)&=& \mbox{Re}\left\{\mathbf{B}_1(\mathbf{r})\exp(i\omega t)\right\}\\
\mathbf{M}_{lab}(\mathbf{r},t)&=& \mbox{Re}\left\{\mathbf{M(\mathbf{r})}\exp(i\omega t)\right\}\\
\mathcal{E}_{lab}(\mathbf{r},t)&=& \mbox{Re}\left\{\mathcal{E}(\mathbf{r})\exp(i\omega t)\right\}.
\end{eqnarray}
For simplicity, and in order to avoid introducing a new notation, we refer to all following field and magnetization quantities as time harmonic quantitites without introducing new symbols.
Using this approach, the reciprocity relationship becomes
\begin{equation}
	\mathcal{E} = \frac{1}{I} i \omega \int_{V} \mathbf{M}(\mathbf{r})\cdot \mathbf{B}_1(\mathbf{r}) \,d \mathbf{r},
\end{equation}
as shown in the Appendix. Here, $\omega$ is the Larmor frequency, $I$ here would indicate the current passed through the receive coil to produce the $\mathbf{B}_1$ field --- it can be simply set to 1 A in order to adjust for the units correctly. It is also important to recognize that the most confusing aspects of reciprocity start from these expressions, not from their derivation. 
Here and in the following, it will be understood that $\mathbf{M}$, and $\mathbf{B}_1$ depend on position, but this will not be written explicitly for brevity of notation.

A further notational simplification will be to consider for now only a single point in space (without taking the integral),  the integral over the whole sample volume can be taken at the end without loss of generality. 

Thus, the starting point for the following discussion will hence be the time-harmonic expression
 \begin{equation}
	\mathcal{E} = c \, \mathbf{M}\cdot \mathbf{B}_1,
	\label{eq:reciproc-harmonic}
\end{equation}
with the proportionality constant being $c=\frac{1}{I} i \omega$. As a consequence of considering a single point in space, the units for the $\mathcal{E}$ here would also be V/m$^3$. Again, we refrain from using a new symbol or differential quantities for the benefit of readability.     

\begin{figure}
\begin{centering}
\includegraphics[width=8cm]{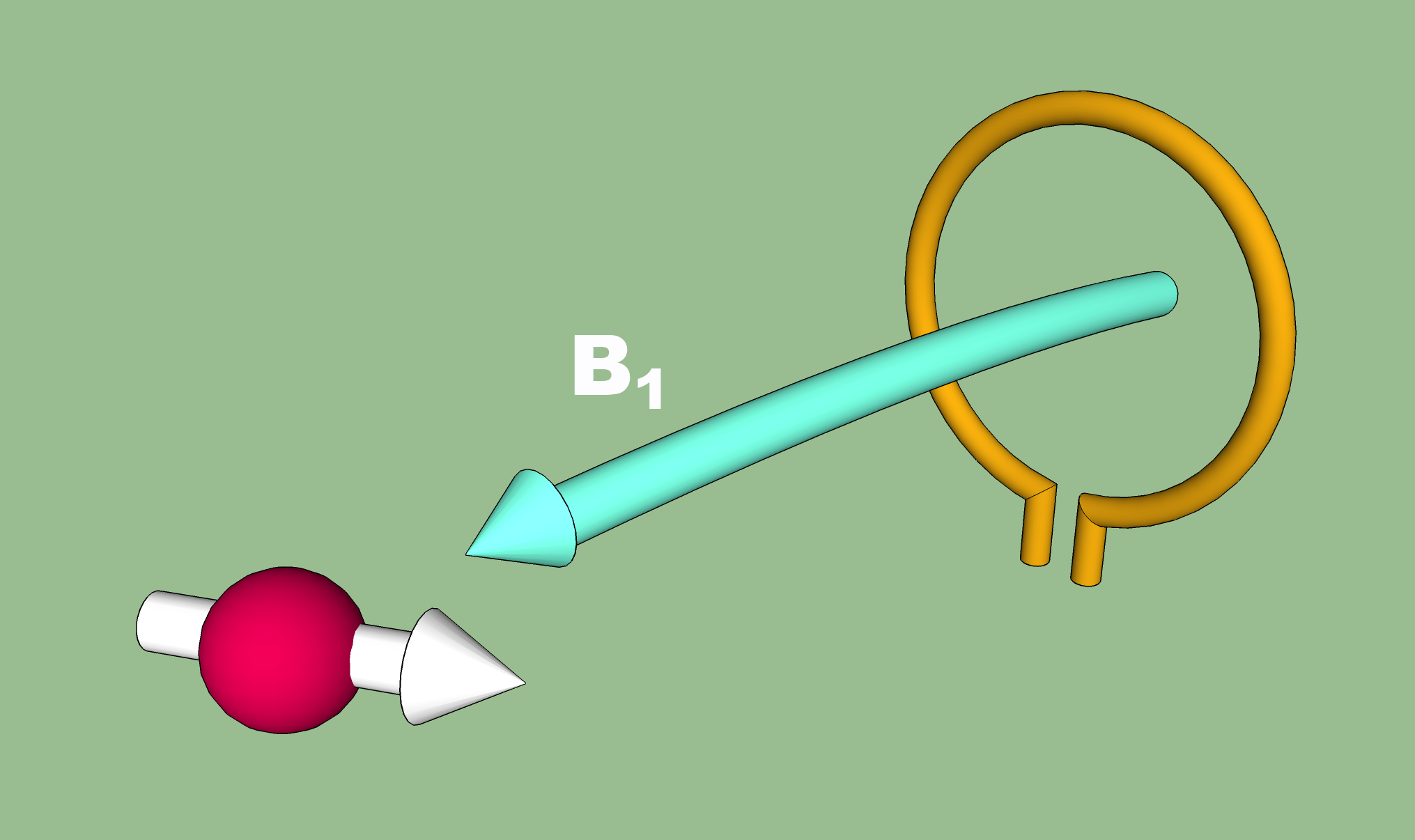}
\par\end{centering}
\caption{\label{fig:illustration}
A spin in the presence of an rf-field generated by a coil. The reciprocity principle provides a relationship between the transmitted and the detected fields.}
\end{figure}

The most useful aspect of this expression for NMR purposes is obtained when one uses for $\mathbf{M}$ the magnetization produced as a result of the application of the rf field, which is further discussed below. 
Let's assume that we apply an rf field of the form 
\begin{equation}
      \mathbf{B}_{1,lab}=C_{x} \cos\left(\omega t + \alpha\right) \mathbf{x} + C_{y} \cos\left(\omega t + \beta\right) \mathbf{y},
      \label{eq:b1lab}
\end{equation} 
which is written in terms of position-dependent amplitudes $C_{x}$, $C_{y}$, and phases $\alpha$, and $\beta$.  
The field is modulated with the Larmor frequency $\omega$ and it may be rotated in the $x-y$ plane at a given position (for example, further near the edges of a solenoid coil or a loop coil). There may also be position-dependent phase delays (for example, due to traveling wave, permittivity, or conductivity effects), which give rise to the phase angles $\alpha$ and $\beta$.  These angles would also describe the situation where one may choose to delay one field with respect to the other, e.g. in a quadrature coil, for example, which will be discussed later. 

The phasor components $B_{1x}$ and $B_{1y}$ can be determined by comparing Eq. (\ref{eq:b1lab}) with
\begin{equation}
      \mathbf{B}_{1,lab}=\mbox{Re}\left[B_{1x} e^{i\omega t} \mathbf{x}\right] 
      + \mbox{Re}\left[B_{1y} e^{i\omega t} \mathbf{y}\right],
      \label{eq:b1lab2}
\end{equation} 
in which case one obtains
\begin{eqnarray}
 B_{1x}&=& C_x e^{i\alpha}\\\nonumber
  B_{1y}&=& C_y e^{i\beta}. \label{eq:defB1x}
\end{eqnarray}
with the phasor field given also as
\begin{equation}
   \mathbf{B}_1= B_{1x}\mathbf{x} + B_{1y}\mathbf{y}
\end{equation}

One can then decompose this expression into co-rotating and counter-rotating field components to obtain
\begin{equation}
      \mathbf{B}_{1,lab}=\frac{1}{2}\left[\mathbf{B}_{1,lab}^+ +\mathbf{B}_{1,lab}^-\right], 
\end{equation}
where
\begin{eqnarray}
       \mathbf{B}_{1,lab}^{\pm}&=&C_{x} \left[\cos\left(\omega t + \alpha\right) \mathbf{x} \pm \sin\left(\omega t + \alpha\right) \mathbf{y}\right] \\
        &&+ C_{y} \left[ \mp \sin\left(\omega t + \beta\right) \mathbf{x}+ \cos\left(\omega t + \beta\right) \mathbf{y}\right],
\end{eqnarray} 
which is equivalent to 
\begin{equation}
      \mathbf{B}_{1,lab}^{\pm}= \mbox{Re}\left[ C_x e^{\pm i (\omega t +\alpha)} (\mathbf{x}- i \mathbf{y}) \right] +  \mbox{Re}\left[ i C_y e^{\pm i (\omega t +\beta)} (\mathbf{x}- i \mathbf{y}) \right].
      \label{eq:corot}
\end{equation}

As is common practice in NMR and related fields, one can safely discount the effect of the counter-rotating field, since it is away from resonance by twice the Larmor frequency. Keeping only the co-rotating component, one therefore obtains
for the phasor $\mathbf{B}_1$ by comparing  Eqs. (\ref{eq:defB1x}) and (\ref{eq:corot})
\begin{equation}
      \mathbf{B}_{1}^+(\mathbf{r})=\left( C_x e^{i \alpha} +i  C_y e^{i \beta} \right)(\mathbf{x}- i \mathbf{y}),
\end{equation} 
and to relate to a  more familiar notation, one could also write
\begin{equation}
      \mathbf{B}_{1}^+(\mathbf{r})=B_{1+} (\mathbf{x}- i \mathbf{y}),\label{eq:B1trans}
\end{equation} 
using the definitions of Eq. (\ref{eq:defB1x}) and $B_{1+}=B_{1x}+iB_{1y}$.

\section{Detected emf after applying a pulse}

For the purposes of this discussion, we will be neglecting relaxation, assume that we have a single resonance at $\omega$, that the rf pulse is applied directly on resonance, and that we are applying the pulse to equilibrium z-magnetization. Furthermore, we do not consider any other internal interactions (e.g. couplings). 

According to the Bloch Equations, such a field will produce a magnetization
\begin{equation}
	M_+ = -i M_0 \sin(\theta) \frac{B_{1+}}{|B_{1+}|},
\end{equation}
where $M_0$ is the equilibrium magnetization, and $\theta=\gamma |B_{1+}| \tau$, with $\tau$ being the pulse duration. The incurred phase of $(-i)$ indicates that the magnetization lags the excitation field by 90$^\circ$.

In vector notation, this expression would be 
\begin{equation}
	\mathbf{M}_+ = -i M_0 \sin(\theta) \frac{B_{1+}}{|B_{1+}|} (\mathbf{x}- i \mathbf{y}).
\end{equation}

At this stage, one may use the reciprocity relation to calculate the detected signal from Eq. (\ref{eq:reciproc-harmonic}) as
\begin{equation}
	\mathcal{E}=c\,\mathbf{B}_{1} \cdot \mathbf{M} = c \, \mathbf{B}_{1} \cdot \mathbf{M}_+, 
\end{equation}
with the second part indicating that the magnetization now contains only one rotating component. Plugging in the expression for 
$\mathbf{M}_+$ and $\mathbf{B}_1=B_{1x} \mathbf{x}+B_{1y} \mathbf{y}$, one obtains
\begin{equation}
	\mathcal{E}= -i c M_0 \sin(\theta) \frac{B_{1+}}{|B_{1+}|}(B_{1x} \mathbf{x}+B_{1y} \mathbf{y})\cdot(\mathbf{x}- i \mathbf{y}) =  -i  c M_0 \sin(\theta) \frac{B_{1+}B_{1-}}{|B_{1+}|}
	\label{eq:emflast}
\end{equation}

This equation suggests that any position- or delay-dependent phases in the $B_1$ field would be eliminated by the scalar product, which is, however, not the case, because $B_{1-}=B_{1x}-iB_{1y}=C_{x}e^{i \alpha}-iC_{y}e^{i \beta}$ (the exponentials are not conjugated), and thus 
\begin{equation}
B_{1+}B_{1-}=(C_{x}e^{i \alpha}+iC_{y}e^{i \beta})(C_{x}e^{i \alpha}-iC_{y}e^{i \beta})=C_{x}^2e^{i 2\alpha}+C_{y}^2e^{i 2\beta},
\label{eq:Cxexpr}
\end{equation}
with the result of phase doubling in $\alpha$ and $\beta$. This effect is also in line with Eq. (28) of Ref. \cite{hoult2000principle}.  We examine this situation in more detail in a couple of examples below. 

If both the coil and the spins were physically rotated by the same angle around the static magnetic field (by rotating the whole apparatus, for example), no global phase factor would arise, because the procedure of the scalar product in Eqs. (\ref{eq:emflast}, \ref{eq:reciproc-harmonic}) would eliminate such a phase. 

In the general case, however, say if the coil and the spins were rotated by different amounts, an additional phase would arise in the complex-valued emf $\mathcal{E}$.  Both, relative physical rotation (between coil and magnetization), and time delay in the transmit field can produce a phase, and hence one can compensate one with the other, which is the strategy employed with quadrature coils (see below).  

The property of uncanceled phases lies at the heart of the ability to detect material-induced phases, and is used in electrical property tomography (EPT) \cite{katscher2009determination}, for example, but is also important for the investigation of signals originating from conductors, as will be discussed below.

\section{Propagating waves}

A simple example of the application of Eq. (\ref{eq:emflast}) would be the analysis of an experiment with an rf field propagating through the sample volume along coordinate $x$, in which case it could be expressed as $B_{1+}=B_{1}^{(0)} \exp(-i k x)$, with the wave-vector $k=\omega/v$, with $v=1/\sqrt{\epsilon\mu}$ being the speed of the wave in the medium. In this case $\alpha=\beta=kx$.

Applying the expression from Eq. (\ref{eq:emflast}) for this case produces 
\begin{equation}
	\mathcal{E}= c M_0\sin(\theta) \frac{(B_{1}^{(0)})^2}{|B_{1+}|}   \exp(-2i k x),
\end{equation}
and the phase originally contained in $B_{1+}$ is doubled as noted before! This is in stark contrast with the phase originating from a global rotation, which cancels out. As has been mentioned by Hoult \cite{hoult2000principle}, an intuitive picture of this phenomenon could also be drawn by examining the situation of an rf field propagating through a coaxial cable: in this case it is perfectly acceptable  that the incurred phase in the measured signal has to double, because the transmitted rf field has to propagate along the cable, and the detected signal has to propagate the same way backwards. A more general description of such processes can be obtained using retarded potentials \cite{INSKO1998111,MAHONY1995145}. Specifically, it is found that the reciprocity expression of Eqs. (\ref{eq:reciproc},\ref{eq:emflast}) still holds, if the retarded  $B_1'$ field is used \cite{INSKO1998111}, with the expression
\begin{equation}
\mathbf{B}_1' = \frac{\mu_0}{4\pi}\int e^{i k \mathbf{r}}(1-ikr)\frac{d\mathbf{l} \times \mathbf{r}}{r^3}.
\end{equation}



\section{Using different coils for excitation and detection}

We will now distinguish between transmit and receive fields, so all relevant quantities, such as $B_\pm$, $C_{x,y}$, $\alpha,\beta,\theta$ will receive a superscript of either $t$ for transmit- or  $r$ for receive-related quantities. Following the same procedure as above to derive the emf, but keeping the transmit ($B_{1\perp}^{t}$) and the receive fields  ($B_{1\perp}^{r}$) separate, one obtains for the equivalent of Eq. (\ref{eq:emflast}),
\begin{equation}
	\mathcal{E}=-i  M_0 \sin(\theta^t) \frac{B_{1+}^tB_{1-}^r}{|B_{1+}^t|}
\end{equation}
with $\theta^t=\gamma |B_{1+}^t| \tau$. The equivalent of Eq. (\ref{eq:Cxexpr}) becomes
\begin{equation}
B_{1+}^tB_{1-}^r=C_{x}^tC_x^re^{i (\alpha^t+\alpha^r)}+C_{y}^rC_y^r e^{i (\beta^r+\beta^r)}.
\label{eq:B1pB1mexpr}
\end{equation}

Here again, it is obvious that any propagation-related phases (due to delays) would add with the same sign. 
For a rotation in space, we have for a counter-clockwise rotaiton by $\phi^r$, and $\phi^t$, respectively
\begin{eqnarray}
      B_+^{t,rot}&=&B_+^{t} \exp( i \phi^t)\\\nonumber
      B_-^{r,rot}&=&B_-^{r} \exp(- i \phi^r),\label{eq:receivephase}
\end{eqnarray} 
and if both were reotated by the same angle $\phi=\phi^t=\phi^r$, then the phase would be canceled in the product $B_{1+}^tB_{1-}^r$ of the reciprocity relation (Eqs. (\ref{eq:emflast},\ref{eq:B1pB1mexpr})).


\section{Quadrature coils,  circular polarization}

Quadrature coils are often employed by placing two coils at an angle of $\phi=\pi/2$ with respect to each other (Fig. 2). The transmit phases for the coils in the ideal case are $\phi_{1}^t = 0$, and $\phi_{2}^t = \pi/2$. In order for the $B_1$ fields originating from the two coils to be aligned with each other at all times, the coils can be supplied with time-shifted rf fields such that these (geometrical) phases are compensated exactly. So, for example, if the rf-field supplied to the second coil is delayed in phase by $\pi/2$ (e.g. by a quadrature hybrid), the overall geometric phase difference between the coils is canceled and the field vectors from both coils align with each other at all times in the rotating frame of the spins.

Because the receive phase $\phi_{r}$ enters the equation with the opposite sign (Eqs. (\ref{eq:receivephase})), in receive mode, the situation is exactly reversed. Therefore, to make sure that the signal combines constructively in detection mode, the signal induced in the second coil now has to be advanced in phase by $\pi/2$ relative to the signal of the first coil before combining them. Therefore, for the recorded signal, the overall phase difference between the signals acquired from each coil would be eliminated.

\begin{figure}
\begin{centering}
\includegraphics[width=8cm]{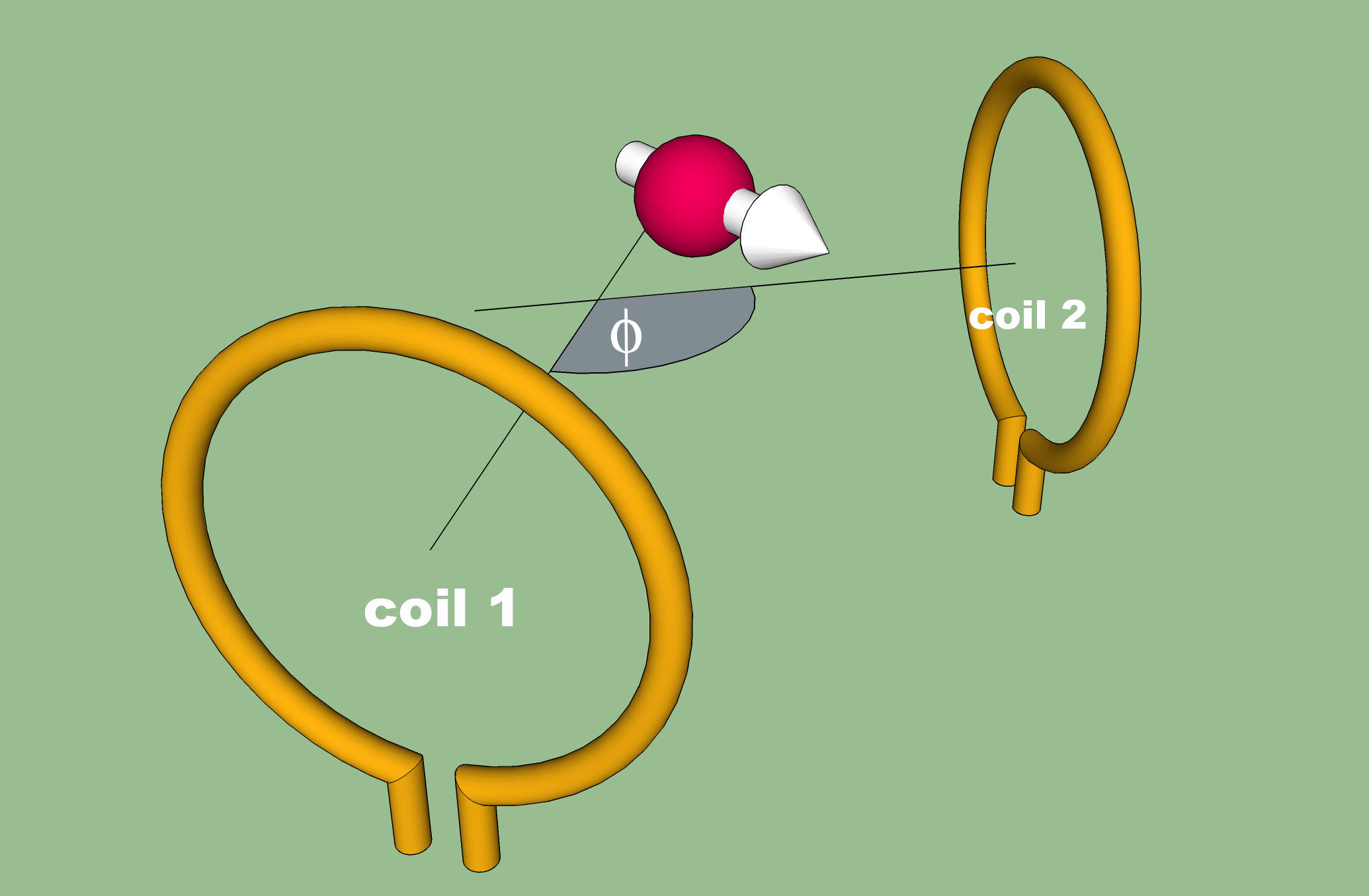}
\par\end{centering}
\caption{\label{fig:2coils}
Two-coil arrangement and quadrature excitation / detection.  The transmit phases of the two coils would be related by $\phi^t_1=\phi^t_2+\phi$.}
\end{figure}

It is clear that this constructive interference in both the transmit and receive mode will not work perfectly over the whole volume but will only be observed in those locations where the geometrical relationship between the two fields is matched by the phase advance (or lag) in the signals supplied to (or received from) the coils. 

Along the same lines, one can analyze a birdcage coil or phased arrays that are supplied with appropriately phase-shifted fields such that the fields they produce combine constructively in the region of interest. In these situations, it is customary to speak of the transmit ``$B_{1+}$" field and the receive ``$B_{1-}$" field. It should be emphasized that the ``$B_{1+}$" and ``$B_{1-}$" labels here refer simply to the way in which the supplied signals are time-shifted with respect to each other in order to compensate for geometry-induced phases and do not represent the actual quantities in the rotating frame! It is more correct to speak of `a coil polarization' mode. This is a point of much confusion, stoked by the use of multiple conventions and notions of the `counter-rotating frame' in the context of reciprocity, because it creates the impression that one should use these quantities directly in Eq. (\ref{eq:emflast}). This approach  would lead to incorrect results. 

For example, in electric property tomography \cite{katscher2009determination}, it is often suggested that one needs to have a good representation of both the transmit $B_{1+}$ and the receive $B_{1-}$ field. While one has direct access to the magnitude of $B_{1+}$ by standard rf field mapping methods, it is unclear how one should estimate $B_{1-}$. Often, the approximation $B_{1+}\approx B_{1-}^*$ is used for lack of direct measures of $B_{1-}$. For example, when attempting to transmit with a ``$B_{1-}$'' polarization, one would enable constructive interference of a rotating field that is opposite to the precession motion of the spins. Therefore, there would be no significant excitation (except for the areas of the sample volume where there are non-ideal phase relationships). By the same token, using a quadrature coil operating in detection mode with ``$B_{1+}$" polarization would again lead to constructive interference of the wrong component, and thus any available signals would  interfere destructively upon combination of the channels.   

The problem of determining transmit and receive fields under these circumstances is rooted in practical and fundamental limitations. If the phase compensation were working perfectly for all locations of interest, there would be no overall transmit or receive phase. One limitation arises from misalignment, spatial distribution of field directions, and the fact that quadrature coils cannot be completely isolated, and a more serious problem arises from the influence of the sample on the fields. Specifically, a significant conductivity in the sample will produce an imbalance between the $B_{1-}$ and $B_{1+}$ fields \cite{vaidya2016dependence}, i.e. $\alpha,\beta \ne 0$ and thus $B_{1-} \ne B_{1+}^*$. Since electrical property tomography aims to measure distributions of conductivities in samples, the very property that one desires to measure produces distortions that complicate such measurements with this imbalance. 

Note, that it may seem that there should be no difference in $B_{1-}$ and $B_{1+}$ fields if a single coil were driven with linear polarization. A sample displaying a significant amount of conductivity would again produce an imbalance between those two fields as described by Vaidya et al. \cite{vaidya2016dependence}. In order to get access to the missing component, one would again need to resort to approximations. 



\section{Detecting a signal from a conducting region}

We now examine the case where the rf field enters a conductive region, and specifically focus on the situation of a good conductor for the purposes of illustrating the phase doubling effect. We follow here closely results presented in Ref. \ref{ilott2014visualizing,ilott2017super}. In such a case, one obtains the expression for the rf field from the Maxwell equations,
\begin{equation}
 \nabla^2 \mathbf{B} = \mu \epsilon \frac{\partial^2 \mathbf{B}}{\partial t^2}+ \mu \sigma \frac{\partial \mathbf{B}}{\partial t}.
\end{equation}
The solution to these equations leads to the plane-wave expression \cite{griffiths1962introduction,jackson1999classical}
\begin{equation}
  B (\mathbf{r}) = B_{10} e^{-\kappa_-  \mathbf{\hat{n}} \cdot \mathbf{r}}  e^{i\kappa_+ \mathbf{\hat{n}} \cdot \mathbf{r} - i\omega t},
 \label{eq:full_field}
\end{equation}
with $B_{10}$ the rf field at the surface of the conductor, $\mathbf{\hat{n}}$, a unit vector, denoting the propagation direction, $\mathbf{r}$ the location vector, and $\kappa_+$ and $\kappa_-$ the real and imaginary parts of the wave vector, $\kappa = \kappa_+ +i\kappa_-$, defined by \cite{jackson1999classical},
\begin{equation}
  \kappa_{\pm}  = \sqrt{\mu\varepsilon} \frac{\omega}{c} \left[ \frac{1}{2} 
  			\sqrt{1 + \left(\frac{2\sigma}{\nu \varepsilon}\right)^2} \pm \frac{1}{2} \right]^{\frac{1}{2}}.
  \label{eq:alpha_beta}
\end{equation}
Here, $\varepsilon$ is the dielectric constant of the conductor and $c$ the speed of light in a vacuum. For a good conductor $\left( \frac{2 \sigma}{\nu \varepsilon}\right) \gg 1$ and $\kappa_+ \approx \kappa_- \approx 1/\delta$ (the inverse of the skin depth constant defined in Eq.~(\ref{eq:delta})), resulting in the same depth-dependence for both the phase and amplitude of the wave. The skin-depth is given as
\begin{equation}
 \delta = \sqrt {\frac{1}{\pi \mu \nu \sigma}},
\label{eq:delta}
\end{equation}
where $\nu$ is the frequency of the field, $\mu$ the permeability of the conductor and $\sigma$ its conductivity.
As an example, for lithium metal, $\sigma=1.08\times 10^7$  S/m and $\varepsilon\approx\varepsilon_0=8.85\times 10^{-12}$ F/m at radio frequencies, defining Li as a good conductor in the frequency regime $\nu \ll  2.44\times 10^{18}$ Hz, and thus well beyond the radio-frequency and microwave regions. 

When incident on a well-conducting surface, assuming that the surface extends to infinity, the boundary conditions dictate that only the rf field parallel to the surface remains, and the field within the conductor in the rotating frame can be described by 
\begin{equation}
 \label{eq:skindepthbp1}
  \tilde B_1(r) = B_{10} e^{-r/\delta}e^{ir/\delta},
\end{equation}
where $r$ denotes the penetration distance from the surface. The rf field decays exponentially and it also acquires a  phase shift linear in the propagation depth, as illustrated in Fig. 3. 

The flip angle imparted on the spin magnetization by this field is given by 
\begin{equation}
	\theta(r)= \gamma \tau \left| B_{10} \right|  e^{-\beta r} = \theta(0) e^{-\beta r},
 \label{eq:flipangle}
\end{equation}
where $\theta(0)$ is the flip angle at the surface of the conductor. 

Using this field in Eq. (\ref{eq:emflast}), and including the volume integral, the voltage induced in the detection coil is given by the integral over the contributions from each depth, 
\begin{equation}
  \mathcal{E} = - i c M_0 \frac{B_{10}^2}{\left| B_{10} \right|}
		\int_{r=0}^{\infty} e^{2i r/\delta}e^{-r/\delta} 
		\sin\left[ \theta(r) \right] dr.
 \label{eq:epscomplex}
\end{equation}
The phase term $e^{2i r/\delta}$ illustrates the aforementioned phase-doubling effect, and governs the extent of constructive or destructive interference between the signals from different depths. An expression equivalent to Eq.~(\ref{eq:epscomplex}) was earlier derived by Mehring et al. \cite{mehring1972influence} and used in NMR/MRI of electrochemical cells and with conducting samples \cite{chandrashekar20127,ilott2014visualizing,ilott2017super}.

\begin{figure}
\begin{centering}
\includegraphics[width=9cm]{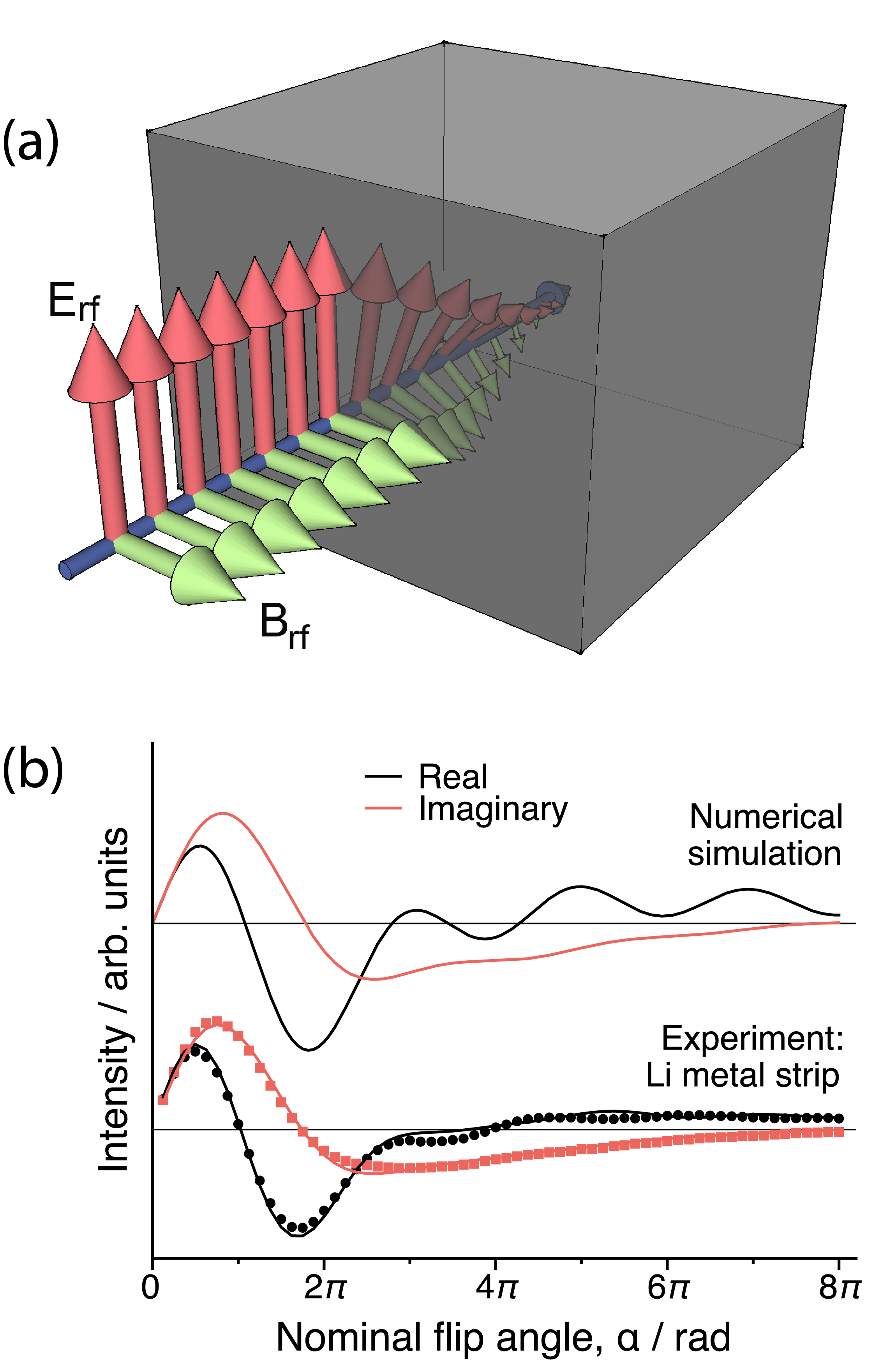}
\par\end{centering}
\caption{\label{fig:penetration}Top: illustration of rf propagation into a conductive region; Bottom: $^7$Li nutation curves of a Li-metal plate sample, along with simulation and experiment. The bottom figure and experimental results were reproduced from \cite{ilott2017super} with permission.}
\end{figure}

A sensitive verification of Eq.~(\ref{eq:epscomplex}) can be produced by a nutation experiment in which the MR signal is measured as a function of the flip angle, $\alpha$,  which is varied experimentally by changing the  pulse duration, $\tau$. An experimental $^7$Li NMR nutation curve performed on a rectangular piece of natural abundance lithium metal (thickness $\gg \delta$) is shown in Fig. 3 along with a numerical simulation of Eq.~(\ref{eq:epscomplex}). 

The nutation curve for lithium metal was obtained on a Bruker Ultrashield 9.4 T Avance I spectrometer operating at 155.5~MHz for $^7$Li, using a Bruker $^1$H$^{7}$Li WB40 birdcage coil for acquisition. The nominal flip angle was calibrated to LiCl(aq) ($\tau_{\pi/2}=38~\mu$s). The sample consisted of a strip of natural abundance lithium metal (Aldrich 99.9\%) cut to ca. 0.4 x 8 x 15~mm and sealed inside a 10~mm NMR tube. Spectra were acquired on resonance with the center of the metal peak and the plotted intensity profiles corresponding to the on-resonance position in the spectrum. 

There is excellent agreement between the experimental curve and the calculated one, particularly at lower flip angles $<3\pi$. Bloch equation simulations including relaxation during the pulse and rf inhomogeneity (20\% variation in $B_{10}$ \cite{ilott2014visualizing}) account for the differences for $\alpha>3\pi$ and produce a good fit with the experimental results.

\section{Conclusions}

We have attempted to provide here a concise summary of the issues encountered in discussions of NMR reciprocity. We specifically avoid the discussion of the negatively rotating frame, which we believe is the source of some confusion, and attempt to delineate other  related topics, such as quadrature detection. One particular example, which provides a good experimental illustration of the effect of `phase doubling' is shown here as well, the case of signals excited and obtained from within a conductive sample region. It is hoped that this article could contribute to clarifications of different aspects of the use of the NMR reciprocity principle. 

\section*{Appendix}

\subsection*{Derivation of reciprocity principle}

We show here one particular derivation of the reciprocity principle. This one is based on Ref. \cite{keun2014electro}.

The fields used in the expressions below will be the time-harmonic fields, given by the definitions 
\begin{eqnarray}
 \mathbf{H}_{1,lab}&=& \mbox{Re}\left\{\mathbf{H}_1(\mathbf{r})\exp(i\omega t)\right\}\\
 \mathbf{B}_{1,lab}&=& \mbox{Re}\left\{\mathbf{B}_1(\mathbf{r})\exp(i\omega t)\right\}\\
 \mathbf{M_{lab}}&=& \mbox{Re}\left\{\mathbf{M(r)}\exp(i\omega t)\right\}\\
 \mathbf{E_{lab}}&=& \mbox{Re}\left\{\mathbf{E(r)}\exp(i\omega t)\right\}.
\end{eqnarray}

 The fields generated by the magnetization $\mathbf{M}$ will be labeled $\mathbf{H}_1^{\mathbf{M}}$, $\mathbf{B}_1^{\mathbf{M}}$, and $\mathbf{E}^{\mathbf{M}}$, and the expression $\kappa=\sigma+i\omega\epsilon$ will be used. With these definitions, we have for the relevant Maxwell Equations in the space without magnetization
\begin{eqnarray}
  \nabla \times \mathbf{E} &=& -i\omega\mu_0 \mathbf{H}_1\\
  \nabla \times \mathbf{H}_1 &=& \kappa  \mathbf{E}+ \mathbf{J}_c,
  \end{eqnarray}
(including a contribution from a filamentary current flow $\mathbf{J}_c$), and
\begin{eqnarray}
  \nabla \times \mathbf{E}^{\mathbf{M}} &=& -i\omega\mu_0 ( \mathbf{H}_1^{\mathbf{M}}+\mathbf{M})\\  
    \nabla \times \mathbf{H}_1^{\mathbf{M}} &=& \kappa \mathbf{E}^{\mathbf{M}},
\end{eqnarray}
in the space where magnetization is present, but without current.

To derive the reciprocity expression, one can start by calculating the emf induced by \textit{unit} current. The induced emf (also time-harmonic here) is then given as follows: 
\begin{eqnarray}
\mathcal{E} =  \oint \mathbf{E}^{\mathbf{M}} \cdot d\mathbf{l}_r &=& \int_{\mathbb{R}^3} \frac{1}{I} \mathbf{E}^{\mathbf{M}} \cdot \mathbf{J}_c d\mathbf{r}\\
 &=& \frac{1}{I}\int_{\mathbb{R}^3} \mathbf{E}^{\mathbf{M}} \cdot (\nabla \times \mathbf{H}_1 - \kappa \mathbf{E}) d\mathbf{r}\\
 &=& \frac{1}{I}\int_{\mathbb{R}^3} \left[ \mathbf{E}^{\mathbf{M}} \cdot \nabla \times \mathbf{H}_1 - \underbrace{\nabla \times \mathbf{H}_1^\mathbf{M}}_{\kappa\mathbf{E}^\mathbf{M}} \cdot \mathbf{E}\right] d\mathbf{r}\\
 &=& \frac{1}{I}\underbrace{\int_{\mathbb{R}^3} \nabla \cdot \left( \mathbf{E}^{\mathbf{M}} \times \mathbf{H}_1 - \mathbf{E} \times \mathbf{H}_1^\mathbf{M} \right) d\mathbf{r}}_{=0}\\
 & & -\frac{1}{I}\int_{\mathbb{R}^3} \left[ \nabla \times \mathbf{E}^{\mathbf{M}} \cdot \mathbf{H}_1 - \nabla \times \mathbf{E} \cdot \mathbf{H}_1^\mathbf{M}\right] d\mathbf{r}.
 \end{eqnarray}

The integrals on the right-hand-sides go over all space.

The first term in the last expression can be cast into a surface integral by the Divergence theorem,
\begin{equation}
	\int_{\mathbb{R}^3} \nabla \cdot \left( \mathbf{E}^{\mathbf{M}} \times \mathbf{H}_1 - \mathbf{E} \times \mathbf{H}_1^\mathbf{M} \right) d\mathbf{r}=\oint_{\mathbb{R}^3} \left( \mathbf{E}^{\mathbf{M}} \times \mathbf{H}_1 - \mathbf{E} \times \mathbf{H}_1^\mathbf{M} \right)\cdot d\mathbf{a}.
\end{equation}
It is standard practice then to take the surface integral at infinity. Using the fact that the terms under the integral drop off faster than $1/r^2$, one can then assume this term to become zero. 

Continuing with this adjustment, one obtains
\begin{eqnarray}
\mathcal{E} &=& \frac{1}{I}\int_{\mathbb{R}^3} \left[ \underbrace{i\omega\mu_0 (\mathbf{H}_1^\mathbf{M}+\mathbf{M})}_{-\nabla \times \mathbf{E}^\mathbf{M}} \cdot \mathbf{H}_1- \underbrace{i \omega\mu_0 \mathbf{H}_1}_{\nabla \times \mathbf{E}} \cdot \mathbf{H}_1^\mathbf{M}\right] d\mathbf{r}\\
 &=& \frac{1}{I} i \omega\mu_0 \int_{\mathbb{R}^3} \mathbf{M}\cdot \mathbf{H}_1\, d \mathbf{r},\\
 &=& \frac{1}{I} i \omega \int_{\mathbb{R}^3} \mathbf{M}\cdot \mathbf{B}_1 \,d \mathbf{r}.
\end{eqnarray}

Other recommended derivations include those by Haacke \cite{brown2014magnetic}, by Hoult \cite{hoult2000principle}, by James Tropp \cite{tropp2006reciprocity}, and by van der Klink \cite{van2001nmr}. A more general treatment of reciprocity is also given in \cite{landau2013electrodynamics}.

\section*{Acknowledgements}

The authors acknowledge funding from the US National Science Foundation under award CHE-1710046 and partially under CBET 1804723. The authors are grateful to Daniel K. Sodickson for stimulating and illuminating discussions and Tom Barbara for suggestions and critical comments.

\end{document}